\documentclass[twocolumn]{aastex631}
\usepackage{CJK}
\usepackage{multirow}

\newcommand{\lya}{Ly$\alpha$}
\newcommand{\oii}{[O~{\sc ii}]}
\newcommand{\oiii}{[O~{\sc iii}]}
\newcommand{\hb}{H$\beta$}
\defcitealias{vanderwel14}{V14}

\received{Sep 20, 2023}
\revised{Feb 29, 2024}
\accepted{Mar 25, 2024}

\begin{document}
\begin{CJK*}{UTF8}{gbsn}
\title{A Close Look at Ly$\alpha$ Emitters with JWST/NIRCam at $z\approx3.1$
%:\\ Extreme Compactness, Low Mass and Young Age
}

\author[0009-0006-4990-7529]{Yixiao Liu (刘一笑)}
\affiliation{Chinese Academy of Sciences South America Center for Astronomy (CASSACA), National Astronomical Observatories(NAOC),
20A Datun Road, Beijing 100012, China}
\affiliation{School of Astronomy and Space Science, University of Chinese Academy of Sciences, Beijing 101408, China}

\author[0000-0002-7928-416X]{Y. Sophia Dai (戴昱)}
\affiliation{Chinese Academy of Sciences South America Center for Astronomy (CASSACA), National Astronomical Observatories(NAOC),
20A Datun Road, Beijing 100012, China}
\correspondingauthor{Y. Sophia Dai}
\email{ydai@nao.cas.cn}

\author[0000-0003-3735-1931]{Stijn Wuyts}
\affiliation{Department of Physics, University of Bath, Claverton Down, Bath BA2 7AY, UK}

\author[0000-0001-6511-8745]{Jia-Sheng Huang (黄家声)}
\affiliation{Chinese Academy of Sciences South America Center for Astronomy (CASSACA), National Astronomical Observatories(NAOC),
20A Datun Road, Beijing 100012, China}
\affiliation{Harvard-Smithsonian Center for Astrophysics, 60 Garden Street, Cambridge, MA, 02215, USA}

\author[0000-0003-4176-6486]{Linhua Jiang (江林华)}
\affiliation{Kavli Institute for Astronomy and Astrophysics, Peking University, Beijing 100871, People's Republic of China}
\affiliation{Department of Astronomy, School of Physics, Peking University, Beijing 100871, People's Republic of China}

\begin{abstract}
We study 10 spectroscopically confirmed \lya\ emitters (LAEs) at $z\approx3.1$ in the UDS field, 
covered by JWST/NIRCam in the PRIMER program. 
All LAEs are detected in all NIRCam bands from F090W to F444W, 
corresponding to restframe 2200\AA--1.2$\mathrm{\mu m}$. 
Based on morphological analysis of the F200W images, three out of the 10 targets are resolved into pair-like systems with separations of $<0.9''$, and another three show asymmetric structures.
We then construct the spectral energy distributions (SEDs) of these LAEs, 
which show little to no extinction. 
All sources, including the pairs, show similar SED shapes, with a prominent flux excess in the F200W band, corresponding to extremely strong  \oiii+\hb\ emission lines (${\rm EW_{rest}}=740$--$6500\,$\AA). 
The median effective radii, stellar mass, and UV slope of our sample are 0.36$\,$kpc, $3.8\times10^7\,M_\odot$, and --2.48, respectively. 
The average burst age, estimated by stellar mass over star formation rate, is $<40\,$Myr. 
These measurements reveal an intriguing starbursting dwarf galaxy population lying off the extrapolations of the $z \sim 3$ scaling relations to the low-mass end: $\sim 0.7$ dex above the star-forming main sequence, $\sim 0.35$ dex below the mass--size relation, and bluer in the UV slope than typical high-z galaxies at similar UV luminosities. 
We speculate that these numbers may require a larger main sequence scatter or tail in the dwarf galaxy regime towards the starburst outliers.

\end{abstract}

\keywords{galaxies: formation -- galaxies: evolution -- galaxies: high-redshift -- galaxies: stellar content -- galaxies: structure}


\section{Introduction} \label{sec:intro}
\lya\ emitting galaxies, or \lya\ emitters (LAEs) have been an extensively studied population since \cite{p&p1967} first elucidated the significance of \lya\ emission in high-redshift (high-$z$) galaxies. Especially in recent decades, as telescope capabilities rapidly enhanced, more and more details about LAEs have been revealed (see \citealt{ouchi20} for a review). Today, we have a consensus that typical LAEs are young, small, and metal-poor galaxies with \lya\ emission originating and escaping from sites of intense on-going star formation \citep[e.g.,][]{nakajima12, hagen14, hagen16, pucha22}. Galaxies of this type are often considered to be the progenitors of present-day galaxies like our own Milky Way \citep[e.g.,][]{gawiser07, hao18, khostovan19}. They are also the most competitive contributors of reionization photons \citep[e.g.,][]{yajima12, finkelstein19, jiang22, matthee22, simmonds23}. While other studies about the high-$z$ Universe focus more on large and/or bright galaxies, which are easier to find, LAEs provide us with a valuable perspective into early-stage galaxies when the evolution just starts. Moreover, with respect to the resonant nature of \lya\ emission, the prominent \lya\ escape fraction of LAEs is a pivotal clue for interpreting properties of their interstellar medium (ISM), circum-galactic medium (CGM), and HI gas in galaxy formation \citep[e.g.,][]{stediel11, momose14, dijkstra16, leclercq20, claeyssens22, reddy22}.

Though an advantageous tool for studying the high-$z$ Universe, many questions regarding the nature of LAEs are still to be resolved. Besides the identification of strong \lya\ emission, our comprehension about the galaxies themselves is still inadequate to explain how \lya\ is exactly produced and treads through the surrounding medium. 
In that sense, revealing the physical structure and galaxy composition are important steps towards understanding the whole picture of LAEs and their place in galaxy evolution. 
However, due to their low continuum emission and high redshifts, 
individual detections at restframe optical bands are hard to attain. In practice, a stacking analysis is usually performed to constrain the average SED of a large sample \citep[e.g.,][]{gawiser06, gawiser07, finkelstein09, ono10, guaita11, vargas14, hao18, kusakabe18}. With large survey programs aimed at LAE, their sample sizes are now growing notably (e.g., a catalogue with over 50k spectroscopically selected LAEs was assembled by HETDEX, \citealt{hetdex23}). In contrast, an in-depth study on individual sources is still missing.

With the advent of the James Webb Space Telescope (JWST), we are finally equipped to inspect these distant, faint targets using resolved images.
Compared to the previous generation infrared telescope Spitzer, JWST introduces an improvement of about one order of magnitude in both limiting flux and spatial resolution. Its preeminent power has been confirmed by a multitude of achievements in searching high-$z$ galaxies within its first year in service. 
Yet most of the current research about LAEs focuses on higher redshifts (e.g, \citet{ning23} at $z\simeq6$; \citet{tang23} at $z=7$--$9$; \citet{jung23} at $z\sim8$; \citet{whitler23} at $z\sim9$; \citet{tang23} at $z=7$--$9$). These works investigate the \lya\ escape process in the reionization era, while paying relatively less attention to the intrinsic properties of the galaxies themselves. 
Where galaxy morphologies and physical properties are considered, the findings conform to previous expectations.
One notable finding is that these LAEs exhibit extremely strong optical line emission, as revealed by both spectral and broadband analyses, which also occurs in the lower redshift ($z \approx 3.1$) sample explored in this paper. Indeed, extreme emission-line galaxies (EELGs) selected by their strong optical line emission \citep[e.g.,][]{vanderwel11,tang21,withers23} show similar properties as LAEs in terms of their size, mass and age, and also feature possible \lya\ emission. This opens a new way of selecting this type of galaxies in the future.

In this paper, we zoom in on 10 LAEs at $z \approx 3.1$ with deep JWST/NIRCam coverage. They were all narrow-band selected and then spectroscopically confirmed by high-resolution spectrographs,
making our sample entirely reliable. 
Their relatively low redshift (compared to the most distant sources now within reach of JWST) eases the observation in terms of signal-to-noise ratio (S/N), and permits NIRCam's observational coverage to sample the rest-frame NIR ($\sim 1 \mu$m) of the low-mass LAE population.
These sources at $z\sim3$ offer a probe of the star formation process in dwarf galaxies entering cosmic noon, and a link between star-forming (SF) galaxies in the local and high-$z$ Universe. 

Our paper is structured as follows.  We briefly introduce our LAE sample and the relevant observations in \S\ref{sec:data}. 
We next describe the morphological and SED analysis in \S\ref{sec:properties}. We investigate the LAE properties in the context of the star-forming main sequence (SFMS) and mass--size relation in \S\ref{sec:results}, discuss their numbers in \S\ref{sec:discussion}, and finally conclude in \S\ref{sec:summary}.

Throughout, we adopt a $\Lambda-$dominated flat cosmology with $\mathrm{H_0 = 70\, km\, s^{-1}\, Mpc^{-1}}$, $\mathrm{\Omega_m=0.3}$ and $\mathrm{\Omega_\Lambda=0.7}$. All magnitudes quoted are in the AB system.

\section{Sample and Data} \label{sec:data}
Our sample is drawn from a large spectroscopically-confirmed LAE sample
at $z\approx3.1$, partly covered by the recently released JWST data from the Public Release IMaging for Extragalactic Research (PRIMER, \citealt{primer-proposal}, PI: J. Dunlop) survey in the UDS field. 

\begin{figure*}
\centering
\includegraphics[width=0.96\textwidth,trim=100 220 100 260,clip]{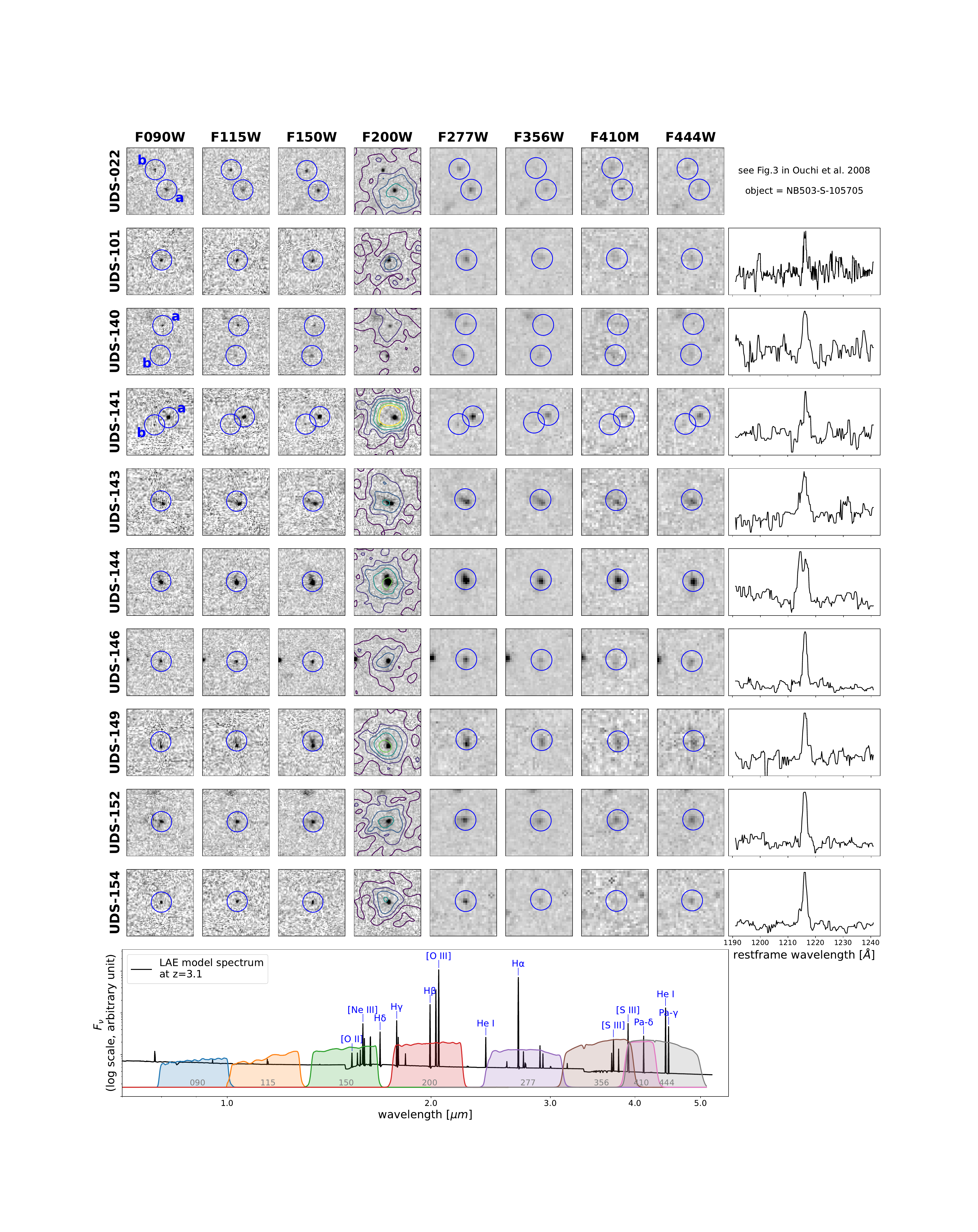}
\caption{Cutouts of the 10 LAE targets in 8 NIRCam bands and their spectra of \lya\ emission. All images are $2\,''\times2\,''$ in size. Blue circles represent apertures with $\mathrm{radii=0.3''}$. 
Overlaid on the F200W cutouts are the narrow-band NB497 contours of 1, 3, 5, 8, 12, 15\,$\sigma$ (except UDS-154, whose NB503 is used for higher S/N).
For the three targets that show up as pair-like systems, we use labels `a' and `b' to mark the major and secondary components respectively.  Consistent IDs will be used in the following sections.
The angular separations between the two components are $0.69, 0.89, 0.47''$ for UDS-022, 140, and 141, respectively. Notice that the orientation of images does not comply with north-south direction. Unsmoothed \lya\ spectra of 9 targets are displayed in the rightmost column. All spectra were observed by M2FS except UDS-101 using Hectospec. The y-axes of spectral panels are in arbitrary flux units. The \lya\ spectrum of UDS-022 with a subtle double-peak feature is shown in Figure 3 of \citet{ouchi08}. In the bottom panel, we display a LAE model spectrum (which is the best fitted model of UDS-144 in SED fitting, see \S\ref{subsec:sed}) along with throughput curves of the 8 NIRCam filters to illustrate the flux contribution in each band. Clearly, emission lines can take a dominant position in the optical-to-NIR spectrum of LAEs.}
\label{fig:all_cutouts}
\end{figure*}

\subsection{LAE Parent Sample} \label{subsec:sample}
Our parent sample includes 166 spectroscopically-confirmed LAEs presented in \citet{guo20}, 
in which 150 targets were observed with Hectospec \citep{fabric05} on the Multi-Mirror Telescope (MMT) or the Michigan/Magellan Fiber System (M2FS; \citealt{mateo12}) on Magellan,
and the remaining 16 are from \citet{ouchi08} observed with the
Faint Object Camera and Spectrograph (FOCAS; \citealt{kashikawa02}) on Subaru and the Visible Multi-Object Spectrograph (VIMOS; \citealt{lefevre03}) on VLT. 
In brief, this parent sample is spread across the wider Subaru XMM-Newton Deep Survey (SXDS; \citealt{Furu08}) field, 
which has a coverage of $\sim$1.2\,deg$^2$ that fully covers the $\sim$150\,arcmin$^2$ 
UDS field observed by JWST. 

The sample was first selected as LAE candidates at $z\approx3.1$ based on narrow-band color excess using narrow-band images taken with the Subaru NB497 (central $\lambda=4986\,$\AA, $\rm FWHM=78\,$\AA) and NB503 (central $\lambda=5030\,$\AA, $\rm FWHM=74\,$\AA) filters.
Sources with a \lya\ equivalent width (EW) $\gtrsim 45\,$\AA\ were selected by requiring a strong narrow-band detection ($>9\,\sigma$ for NB503 or $10\,\sigma$ for NB497, due to different depths) and the dropout of continuum flux (BV$\,-\,$NB$\,{\rm >1\,mag}$, where BV is the weighted magnitude between B and V bands, and NB is the magnitude in NB497 or NB503).

These candidates were then observed with different spectrographs at resolutions ranging from $R\sim 500$--$2000$.
All possible low-z contaminations (e.g. \oii, \oiii, H$\alpha$) were carefully removed, 
before the confirmation of LAE systems.
The restframe EW of the parent LAE sample has a median value of $\sim 90\,$\AA. 
The redshifts determined by the peak of the \lya\ line are within the range $z=3.05$--$3.16$.  

We then matched the parent LAE sample with the PRIMER released data in the UDS field, 
and found 10 spectroscopically confirmed LAEs, 
including three possible pair systems (see \S\ref{subsec:morphology}) 
that have been observed by JWST. 
Of the 10 LAEs in our sample, 
9 are confirmed by \citet{guo20} and one by \citet{ouchi08}. 
The available \lya\ spectra of the 9 LAEs from \citet{guo20},
observed with either Magellan M2FS or Hectospec,
are shown in Figure~\ref{fig:all_cutouts}.
We note that our 10 objects happen to be relatively faint among the parent sample, 
with a $V$-band magnitude in the range 25.6--26.8, 
while the parent sample has a $V$-band magnitude range spanning ${\rm mag}_V=23.4$--$27.4$.

\subsection{JWST Data} \label{subsec:data}
\subsubsection{Observation and Data Reduction} \label{ssubsec:reduction}
The PRIMER project surveys the well-observed UDS and COSMOS fields 
with JWST NIRCam and MIRI in a parallel mode. 
For NIRCam imaging, 8 filters are used, 
including F090W, F115W, F150W, F200W with the Short Wavelength (SW) detectors,
and F277W, F356W, F410M, F444W with the Long Wavelength (LW) detectors. 
The first two epochs of UDS observations 
were completed in July 2022 and January 2023, respectively. 

We retrieve the \texttt{`.uncal'} data that overlap with our parent sample coverage from the Mikulski Archive for Space Telescopes (MAST)\footnote{\url{https://mast.stsci.edu/}}.
Then we reduce the data using the official \texttt{JWST Pipeline (v1.9.4)}. 
The calibration reference file involved in the reduction is \texttt{`jwst$\_$1046.pmap'}. 
Two additional processes called `snowball' and `1/f noise' are 
added in Stage-1 and after Stage-2, respectively, to complement the current pipeline. 
We note that some exposures %\textcolor{red}{(\#XXXXX)}
of the UDS field are contaminated by unknown stripes in the NIRCam data.
According to the JWST Help Desk \footnote{\url{https://stsci.service-now.com/jwsts}}, 
this contamination may be caused by the scattered light from some surface or optical path 
given its wavelength dependence. 
To mitigate its effect, 
we implement an additional step to first measure and then deduct 
the median value along the stripes during the 1/f reduction process. 
The final images are drizzled to a uniform scale of 0.03$''$/pixel for the SW and 0.06$''$/pixel for the LW data, respectively. 
We notice that the WCS of the SW and LW filters 
often shows an offset,
due to the paucity of reference stars in the field of view.
To correct this, we adopt the F200W image as the SW reference, 
and apply to the LW images an average WCS shift,
generated by matching the catalogs of each filter to the F200W catalog. 
Figure~\ref{fig:all_cutouts} displays the postage stamp images of our 10 targets with a $2\,''\times\,2\,''$ size. 
Since some of the fields are covered by two pointings, the total exposure time for each source varies, as summarized in Table~\ref{tab:info}.

\subsubsection{Photometry}
\label{subsec:photometry}
We measure the fluxes of our objects within circular apertures. 
Due to their compact sizes (see \S\ref{subsec:size}), 
either a $r=0.1''$ or $r=0.3''$ aperture is chosen for the aperture flux measurement.
The choice is made by comparing the flux values after PSF correction for different apertures (0.1--0.4$''$),
and when the measured fluxes are consistent between two apertures, 
the smaller aperture, typically also associated with smaller uncertainties, 
is adopted.
In general, if our target is compact with no extended structure or irregular shapes, 
a $r=0.1''$ aperture is adopted, 
while for sources with possible extended features,
aperture of $0.3''$ is often adopted. 
For the same object, the same aperture is chosen for all bands, and with corresponding 
PSF correction factors\footnote{https://jwst-docs.stsci.edu/jwst-near-infrared-camera/nircam-performance/nircam-point-spread-functions} to reach the total flux. 
Our photometry results and the choice of apertures are listed in Table~\ref{tab:info}. 
We apply a local background measurement and removal in an area of $3\,''\times3\,''$, 
after masking out all pixels with $>3\,\sigma$ values.
The flux errors are computed by $\sqrt{n_{pix}}\,\times\,\mathrm{ERR_{max}}$, 
where $n_{pix}$ is the number of pixels inside the aperture and $\mathrm{ERR_{max}}$ is the maximum error value within the aperture. 
Additionally, to take potential systematic uncertainties into account, we set a lower error limit of 10\% of the total flux. 
Table~\ref{tab:info} lists the measured photometry for all sources in our sample.

\subsection{Ancillary Data}\label{subsec:ancillary}
The UDS and SXDS fields are rich in ancillary data,
covering a wide wavelength range from the X-ray to the submillimeter. 
For the following analysis, we also include the photometry in 6 other bands,
including the CHFT-MegaCam/$u$ and the Subaru-SupCam/$B,V,R_c,i',z'$ to complement the observations.  These data are taken from the catalogue by \citet{galametz13}.
Given the low resolution of the UV-optical data (resolution $\sim 0.8''$), 
their photometry was measured by template-fitting with prior segmentation information based on the 
CANDELS WFC3 F160W image, whose resolution is $\rm FWHM=0.2''$ and limiting magnitude 27.45 mag.
measured by template-fitting method using the source detection in F160W as reference. 
Among the 10 LAEs in our sample, 7 are recorded in their catalogue. 
Two (UDS-152, 154) of the remaining three did not have F160W coverage,
while the other one (UDS-140) was covered but not detected. 
With our new NIRCam bands, a total of 14 bands are included in the following SED analysis (see \S\ref{subsec:sed}), spanning from the restframe UV to NIR, 
allowing stellar mass, UV slope and SFR measurements. 

Though active galactic nuclei (AGNs) are very rare in dwarf galaxies such as those included in our sample, we do as a sanity check examine the Spitzer/MIPS 24$\mu$m, 
corresponding to rest-frame near-IR 6$\mu$m,  
and Chandra full-band images in the UDS field for possible contamination from strong AGNs.
The resolution of these images is $6''$ (24$\mu$m) and 0.2--0.7$''$ (Chandra), respectively. 
We do not find any detections in the catalog or by visual inspections in these bands. 
Nor did we see any radio or submillimeter counterparts at their locations \citep{heywood20, geach17}.
Therefore, in the following analysis, we assume no AGN contribution to our sample.

\movetabledown=2.5cm
\begin{rotatetable}
\begin{deluxetable*}{lccccccccccc}
\centerwidetable
\tabletypesize{\small}
\tablecaption{Effective Radii and Photometry for 10 LAEs and Their Possible Companions at $z\simeq3.1$ \label{tab:info}}
\tablehead{
\colhead{ID}  & \colhead{Exp. time}  & \colhead{$R_{\mathrm{e,circ}}$}  & \colhead{F090W}  & \colhead{F115W}  & \colhead{F150W}  & \colhead{F200W}  & \colhead{F277W}  & \colhead{F356W}  & \colhead{F410M}  & \colhead{F444W}\\
\colhead{}  & \colhead{(s)}  & \colhead{(kpc)}  & \colhead{(mag)}  & \colhead{(mag)}  & \colhead{(mag)}  & \colhead{(mag)}  & \colhead{(mag)}  & \colhead{(mag)}  & \colhead{(mag)}  & \colhead{(mag)} \\
\colhead{(1)} & \colhead{(2)} & \colhead{(3)} & \colhead{(4)} & \colhead{(5)} & \colhead{(6)} & \colhead{(7)} & \colhead{(8)} & \colhead{(9)} & \colhead{(10)} & \colhead{(11)}
}
\startdata
UDS-022a  	&1675.0 	&$0.36\pm0.04$ 	& 27.32$\pm$0.11 	& 27.62$\pm$0.12 	& 27.35$\pm$0.10 	& 26.89$\pm$0.10 	& 27.47$\pm$0.10 	& 28.21$\pm$0.12 	& 27.88$\pm$0.18 	& 28.18$\pm$0.20 \\
UDS-022b  	&1675.0 	&$0.21\pm0.08$ 	& 28.03$\pm$0.19 	& 28.17$\pm$0.20 	& 27.73$\pm$0.12 	& 27.35$\pm$0.10 	& 27.84$\pm$0.10 	& 29.16$\pm$0.27 	& 27.84$\pm$0.17 	& 28.20$\pm$0.20 \\
\multirow{2}*{UDS-101\tablenotemark{$\ast$}} & 837.5(SW)  &  \multirow{2}*{$0.53\pm0.12$} & \multirow{2}*{27.13$\pm$0.34} & \multirow{2}*{26.73$\pm$0.21} & \multirow{2}*{26.72$\pm$0.16} & \multirow{2}*{26.43$\pm$0.11} & \multirow{2}*{26.96$\pm$0.10} & \multirow{2}*{27.43$\pm$0.14} & \multirow{2}*{28.50$\pm$0.58} & \multirow{2}*{27.66$\pm$0.25} \\
& 1675.0(LW) & & & & & & & & & &\\
UDS-140a  	&1675.0 	&$0.23\pm0.13$ 	& 28.42$\pm$0.32 	& 27.92$\pm$0.19 	& 28.73$\pm$0.30 	& 27.82$\pm$0.12 	& 28.13$\pm$0.11 	& 29.42$\pm$0.32 	& 28.57$\pm$0.30 	& 28.92$\pm$0.34 \\
UDS-140b  	&1675.0 	&$0.38\pm0.21$ 	& 28.21$\pm$0.25 	& 28.61$\pm$0.32 	& 28.18$\pm$0.19 	& 27.52$\pm$0.10 	& 28.11$\pm$0.11 	& 28.17$\pm$0.11 	& 27.83$\pm$0.17 	& 28.11$\pm$0.17 \\
UDS-141a  	&837.5 	&$0.22\pm0.03$ 	& 26.34$\pm$0.10 	& 26.52$\pm$0.10 	& 26.50$\pm$0.10 	& 25.95$\pm$0.10 	& 26.70$\pm$0.10 	& 27.36$\pm$0.10 	& 27.18$\pm$0.14 	& 27.18$\pm$0.11 \\
UDS-141b  	&837.5 	&$0.05\pm0.33$ 	& 28.86$\pm$0.60 	& 29.35$\pm$0.78 	& 28.94$\pm$0.49 	& 27.59$\pm$0.16 	& 28.47$\pm$0.21 	& 28.84$\pm$0.28 	& 28.59$\pm$0.42 	& 28.54$\pm$0.35 \\
UDS-143 \tablenotemark{$\ast$}
 	&837.5 	&$0.55\pm0.05$ 	& 26.55$\pm$0.21 	& 26.25$\pm$0.18 	& 26.11$\pm$0.11 	& 25.79$\pm$0.10 	& 26.17$\pm$0.10 	& 26.32$\pm$0.10 	& 26.34$\pm$0.14 	& 26.30$\pm$0.10 \\
UDS-144 \tablenotemark{$\ast$}
 	&1675.0 	&$0.44\pm0.02$ 	& 25.86$\pm$0.10 	& 25.82$\pm$0.10 	& 25.44$\pm$0.10 	& 24.75$\pm$0.10 	& 25.26$\pm$0.10 	& 25.86$\pm$0.10 	& 25.61$\pm$0.10 	& 25.66$\pm$0.10 \\
UDS-146 \tablenotemark{$\ast$}
 	&1675.0 	&$0.27\pm0.03$ 	& 26.84$\pm$0.16 	& 27.16$\pm$0.21 	& 27.03$\pm$0.15 	& 26.09$\pm$0.10 	& 26.62$\pm$0.10 	& 27.10$\pm$0.10 	& 27.64$\pm$0.26 	& 27.14$\pm$0.16 \\
UDS-149 \tablenotemark{$\ast$}
 	&837.5 	&$0.43\pm0.08$ 	& 26.60$\pm$0.20 	& 26.67$\pm$0.20 	& 25.77$\pm$0.10 	& 25.74$\pm$0.10 	& 26.15$\pm$0.10 	& 26.75$\pm$0.11 	& 26.41$\pm$0.14 	& 26.48$\pm$0.13 \\
UDS-152 \tablenotemark{$\ast$}
 	&1675.0 	&$0.98\pm0.50$ 	& 26.57$\pm$0.12 	& 26.58$\pm$0.12 	& 26.46$\pm$0.10 	& 25.93$\pm$0.10 	& 26.15$\pm$0.10 	& 26.57$\pm$0.10 	& 26.65$\pm$0.12 	& 26.36$\pm$0.10 \\
UDS-154  	&837.5 	&$0.19\pm0.03$ 	& 27.51$\pm$0.21 	& 27.34$\pm$0.17 	& 27.17$\pm$0.13 	& 26.12$\pm$0.10 	& 27.10$\pm$0.10 	& 28.18$\pm$0.17 	& 28.21$\pm$0.29 	& 27.65$\pm$0.17 \\
\enddata
\tablecomments{
(1) Object ID. (2) Exposure time in seconds. Different bands share the same exposure time except in the case of UDS-101, which falls in the gap between short wavelength detectors (SW) in one pointing, resulting in a shorter exposure time. (3) Circularized $R_e$ in F200W band derived from GALFIT, which is calculated by the effective radius of semi-major axis times $\sqrt{b/a}$. The angular-diameter distance is calculated based on spec-z and the scale at $z\approx3.1$ is $\mathrm{\sim7.6\,kpc}$ per arcsec.  (4)--(11) Photometry in 8 NIRCam filters.}
\tablenotetext{\ast}{For these relatively large and bright sources, we use r=0.3" apertures to measure their flux. Different PSF correction factors are adopted depending on the aperture sizes.}
\end{deluxetable*}
\end{rotatetable}
\clearpage

\section{Galaxy Properties} \label{sec:properties}
In this section, we introduce the measurement of physical properties of sources in our sample including their morphology, physical size, and properties like stellar mass, and SFR derived from the SED fitting. Specific methods and details are presented in each subsection. 
\subsection{Morphology} \label{subsec:morphology}
The PSF sizes of NIRCam filters are between 0.03--0.14$''$ from short to long wavelengths. Multiplied by the angular scale at $z\approx3.1$ ($\mathrm{\sim7.6\,kpc/''}$), the NIRCam images allow us to distinguish structures on scales down to $\sim0.3\,$kpc. Though their compact nature does not allow to fully resolve the sources in our sample, we leverage the sharper view than any pre-JWST observation to place robust constraints on their sizes.

As shown in Figure~\ref{fig:all_cutouts}, out of the 10 LAE systems in our sample, 3 exhibit obvious extended (and asymmetric) structure in JWST images (UDS-143, 144, 149), while the other 7 appear as point-like sources. 
Among the point sources, three of them (UDS-022, 140, 141) also show a possible companion with 0.47--0.89$''$ separations. 
We note that a couple other targets (UDS-146, 152) also have additional counterparts at $\sim\,1\,''$ away
(separation from target: 0.98$''$, 0.91$''$, respectively), 
however, given that their separations are larger than the NB resolution ($\sim\,0.8''$),
we do not consider them as pair candidates. 
% given thatthey are resolvable in the NB images (PSF$\sim\,0.8''$) but were not selected as LAE candidates. 
In addition, 
we also only keep the pair candidates when the secondary object shares similar NIRCam SED with the primary,
indicative of genuine physical pair system at similar $z$.
In fact, our photo-$z$ test, though limited to the 8 NIRCAM filters that can resolve multiple objects,
show consistent redshift estimates of the pair members in these 3 candidate systems, 
but not for the two with larger separations.
We verified the 8-filter NIRCam SEDs of these sources and their companions, and confirm the presence of a prominent F200W excess in each of them, similar to what is seen for the other, non-pair LAEs.  This excess is caused by the dominant contribution of optical emission lines (specifically \oiii+\hb) that redshift into this filter (see the bottom panel of Figure~\ref{fig:all_cutouts} for an illustration).  
An imprint of (more modest) line contributions in the adjacent F277W filter is also commonly seen, and attributed to the H$\alpha$ line which redshifts into this band.  
In addition to their similar SED shapes,
the photo-z fitting results using the NIRCAM bands
show comparable redshifts for both members at $z\sim$3.1,
consistent with them being physically associated.
There is no explicit answer to whether the \lya\ emission comes from one or both objects in the 
pair based on the spectroscopy or narrow-band images (Figure~\ref{fig:all_cutouts}, column 4). 
In addition, the double-peaked nature (with peak separation of $\lesssim 3\,$\AA\ in the restframe) of 
the \lya\ emission observed for UDS-022 and 144 
is suggestive of pairs but not conclusive, 
as such line profiles 
can also arise from the complexities of \lya\ radiative transfer \citep[e.g.,][]{verhamme06}.  
What we can conclude with certainty is that a low-z \oii\ doublet interloper interpretation is ruled out on the basis of the SED (and specifically the F200W excess) at high confidence.
Considering the similarity of properties in all aspects, we treat the pair-like sources equally and include the companions as individual sources in later sections. To distinguish objects in pairs, we label the brighter one with the letter `a' and the fainter one with letter `b'.

\subsection{Effective Radii} \label{subsec:size}
The size measurement is conducted with \texttt{GALFIT} code \citep{galfit}. In the fitting, we adopt a single S\'ersic model to fit the target galaxy and use the WebbPSF \citep{webbpsf14} for PSF deconvolution. 
In general, the semi-major axis of the ellipse that contains half of the total flux are chosen as effective radius. 
But in the case that the target galaxy is small, the observed flux distribution is largely dominated by the PSF, which means that changes to the morphological parameters like S\'ersic index or ellipticity would not significantly alter the final model, so that these parameters cannot be well constrained. 
When conducting the  fittings, we find that the best-fitting ellipticity and the corresponding semi-major $R_e$ vary slightly with the initial input values, while the circularized $R_{e,circ}$ (defined as $\sqrt{b/a}\,R_e$, where $b/a$ is the ratio between semi-minor and semi-major axis) is more robust and less affected. 
Therefore in the following discussions, we use the circularized radius $R_{e,circ}$ as the physical sizes of our galaxies. 
As shown in the bottom panel of Figure~\ref{fig:all_cutouts}, the flux of our objects contains strong emission-line contributions in certain bands (see the end of \S\ref{subsec:sed}). 
Since we do not want our size measurement to be biased towards a specific sub-population, we run the \texttt{GALFIT} in 5 bands, F115W, F150W, F200W, F277W, and F356W, respectively. 
We find consistent size measurement in all five bands using \texttt{GALFIT}, 
with $R_{e,circ}$ measured in different bands agreeing within their error margins. 
In the following analysis,
we choose the sizes derived from F200W images (listed in Table~\ref{tab:info}) as the representative size, 
as it lies in a wavelength range well above the Balmer break 
and has the highest imaging quality.

As a sanity check, we also measured the angular sizes of all objects in F200W by fitting Gaussian functions to their 1D flux distribution, despite the asymmetrical morphology of some sources.
Overall, the observed FWHM of these sources are 1.30--2.35 $\times$ the PSF size. 
We deduct the instrumental resolution in quadrature and obtain the physical effective radius $R_e$ through $R_e=\mathrm{0.5\times FWHM_{phy}=0.5}\times d_A\times \mathrm{\sqrt{{FWHM_{obs}}^2-{FWHM_{PSF}}^2}}$, where $R_e$ is the radius that contains half of the total flux in a 2-D Gaussian profile,
and $d_A$ is the angular scale for the redshift of the source. 
The final results agree well with the $R_{e,circ}$ derived from \texttt{GALFIT} except for one outlier, UDS-152, whose \texttt{GALFIT} size is nearly two times larger than the 1-D fitted size. 
The \texttt{GALFIT} value is also distinctly larger than the other sources. In Table~\ref{tab:info} and the following discussion, just as all other sources, we still adopt the \texttt{GALFIT} size of UDS-152 as the center value, but assign a large enough uncertainty to cover the result by the two different methods.

\begin{figure} %[t]
\centering
\includegraphics[width=0.48\textwidth,trim=80 10 80 10,clip]{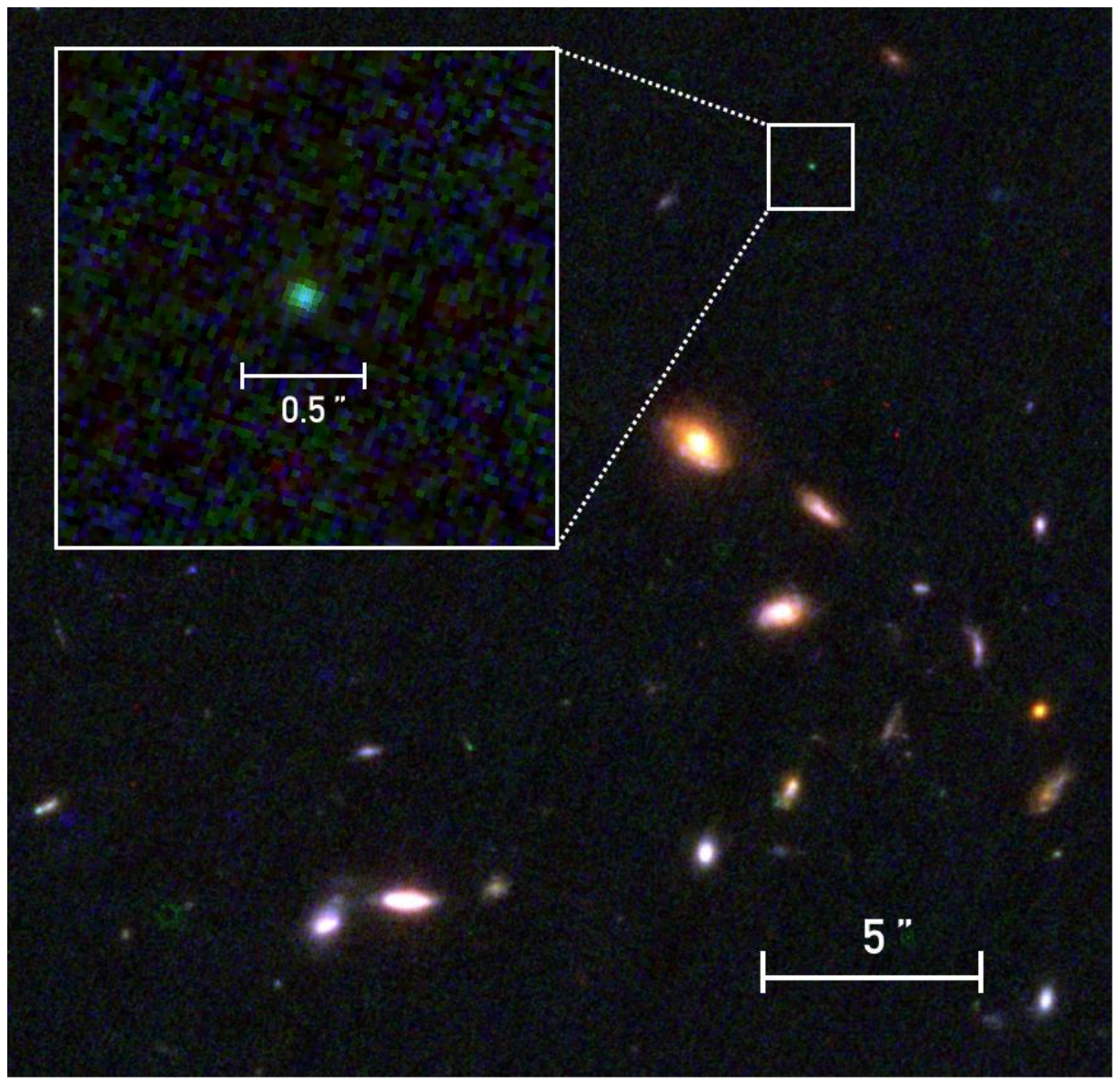}
\caption{A false-color image zooming in on the source UDS-154. The RGB color composite combines F115W (blue), F200W (green) and F356W (red), matching the real color in restframe. Angular scales are indicated both on the original and on the zoomed-in frame. \label{fig:colorimg}}
\end{figure}

The median $R_{e,circ}$ of our sample in F200W band given by \texttt{GALFIT} is 0.36$\,$kpc (0.32$\,$kpc if we exclude UDS-152), which is extremely small for known galaxies (e.g. the median size value for late-type galaxies in SDSS survey with similar restframe $r$-band magnitudes is $\sim2\,$kpc, \citealt{shen03}).
In Figure~\ref{fig:colorimg}, we demonstrate a pseudo color image of the source UDS-154 as an example (R: F356W, G: F200W, B: F115W, approximately corresponding to the restframe visual color for objects at $z\approx3.1$). 
The minuscule size and green color are two salient characteristics of our sources. 
In fact, all of our sources appear blue or green in the color images,
suggesting low dust extinction (see also Section~\ref{subsec:sed}). 
These features are reminiscent of Green Pea galaxies (GPs), which are thought to be 
the local analogs of LAEs \citep[e.g.,][]{cardamone09, malhotra12, kim21, rhoads23}.

\begin{figure} %[t]
\centering
\includegraphics[width=0.43\textwidth,trim=0 58 20 110,clip]{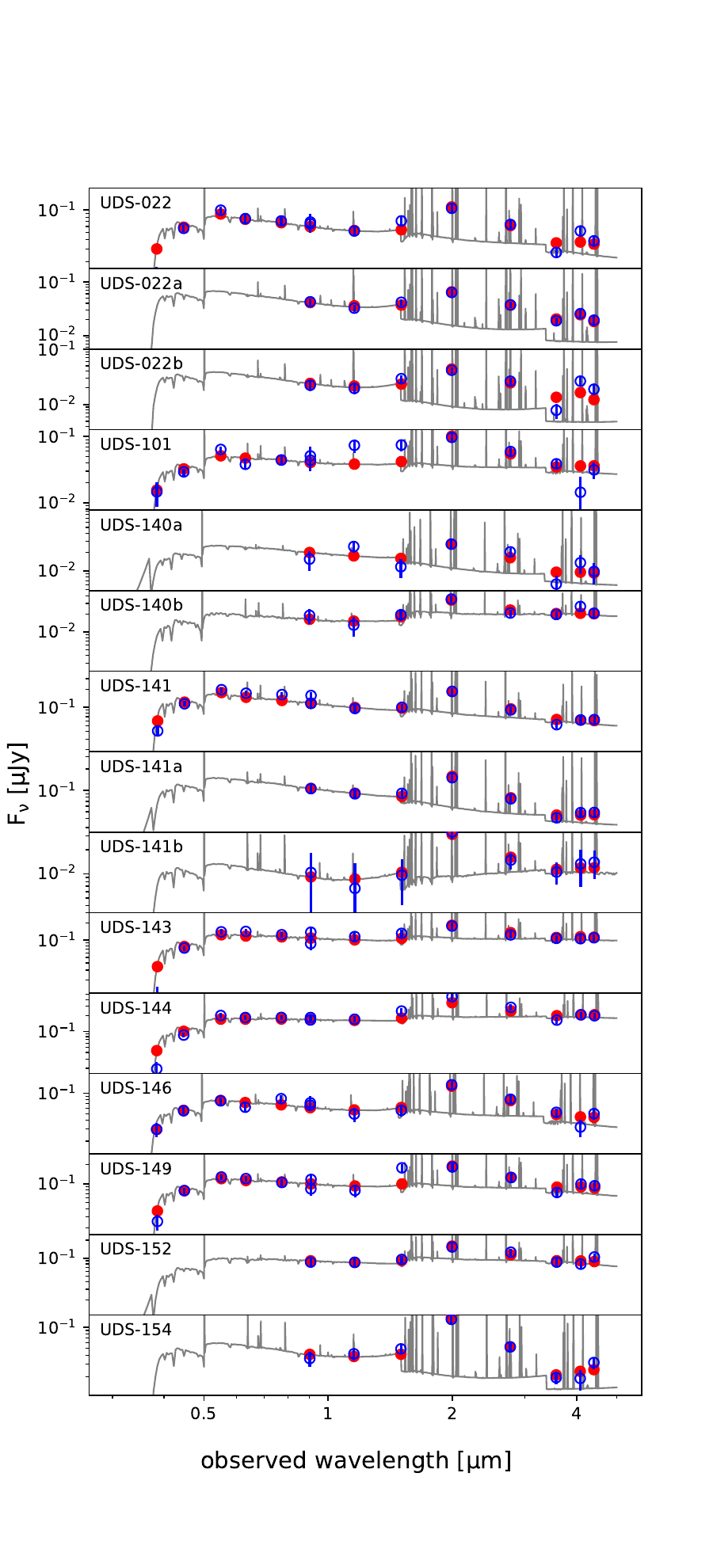}
\caption{Best-fitting model spectra (gray lines), model fluxes (red dots) and observed fluxes (blue circles with error bars). Each panel presents one {\tt CIGALE} fit, with the ID of its fitted target labelled in the upper left corner. The SEDs with 14 observed flux points comprise photometry in CFHT-$u$ + SupCam-$B,V,R_c,i',z'$ + 8 NIRCam bands. Other spectra are fitted using only the 8 NIRCam bands. All fittings are conducted under the same parameter space (see Table~\ref{tab:cigale_para}). For the 3 targets (UDS-022, 140, 141) that have a possible companion within small separations, we fit each of their components individually based on their NIRCam fluxes. However, for UDS-022 and UDS-141, we also have their unresolved $u$-to-$z'$ band photometry, so we carry out two additional fits incorporating the shorter-wavelength photometry and adding the NIRCam fluxes of their two components together (i.e., regarding the pairs as a whole). \label{fig:sed}}
\end{figure}

\subsection{SED Fitting, Stellar mass, and SFR}\label{subsec:sed}
We utilize the stellar population modelling code {\tt CIGALE} \citep{cigale2019} to conduct SED fitting with the redshift fixed to its spectroscopic value. This code allow us to fit SEDs accounting for the estimated nebular emission based on CLOUDY models. The best-fitting templates are displayed in Figure~\ref{fig:sed}. As mentioned in Section~\ref{subsec:ancillary}, we fit 7 targets with 14-band photometry (CHFT-$u$ band + 5 Subaru-SupCam bands + 8 NIRcam bands). For UDS-022 and 141 among these 7 targets, since it is hard to separate their two components in the unresolved ground-based observations, we instead sum the JWST flux of objects `a' and `b' to treat them as an integral whole. For separate sources in pairs and other targets without archival optical photometric data, we fit their SEDs with only 8 NIRCam bands. 
In all cases, we run {\tt CIGALE} with identical models and parameter space as listed in Table~\ref{tab:cigale_para} and described below. We choose a delayed star formation history with a constant burst in the recent 1--50 Myrs. 
A wide parameter space is given to both populations to test various SFHs from single old population to burst-dominated histories. 
For the stellar population, we use the classic \citet[][BC03]{Bruzual2003} models and a \citet{Chabrier2003} stellar initial mass function (IMF). 
Considering the young age and strong emission-line that our sample has, we fix the metallicity of the stellar component to sub-Solar ($Z_*=0.2\,Z_\odot=0.004$). 
In the nebular model, we specify a high range of ionization parameters $\log U$ from --2.2 to --1.2. Satisfactory fits are obtained for all sources, with reduced-$\chi^2\lesssim2$.

We check the consistency of SED fitting with and without the inclusion of optical bands 
by comparing the estimated stellar masses of the same sources.
For the 7 SEDs where this test was feasible, differences never exceeded 0.033 dex, with a mean deviation of 0.007 dex, which is negligible. 
Furthermore, the total masses of the two components `a' and `b' for UDS-022 and 141 agree with the total masses derived directly by modelling the integrated SEDs, within the margin of error. We thus regard these two fitting methods as consistent. 

In Table~\ref{tab:properties}, we list the properties of all 13 separate sources derived by SED fitting. 
The UV slopes of the best-fitting spectra have a range of --2.8 to --2.1 and a median of --2.48.  This shows that all of our Lya targets are very blue with little dust attenuation compared with other high-z SF populations \citep{Bouwens2014, reddy18}. 
The $\rm A_v$ based on SED fitting has a range of 0.05-0.41 (mean value is 0.21), 
indicating little to no dust extinction. 
For comparison, we also adopt the method described by \citet{meurer99},
which yields similarly small values of $A_{1600}=$0-0.15, 
confirming their low level of extinction.
The median stellar mass of our sample equals to $3.78\times10^7\,M_\odot$.

Using the best-fitting models, we then calculate the SFRs based on their uncorrected UV luminosities.\footnote{Attenuation according to the \citet{Calzetti2000} law was allowed in the fitting, but effectively not (or negligibly) used in the best-fitting solutions, such that working with intrinsic (i.e., dust-corrected) SFR values would not alter our conclusions.  By the same token, adoption of a steeper attenuation law as appropriate for e.g. the Small Magellanic Cloud (SMC) would not impact our results either.} The luminosity is averaged among 1500--2800\AA\ and we convert it into SFR using the calibration given by \citet{kennicutt98} under the assumption of a Chabrier IMF:
\begin{equation}\label{eq:sfr}
\mathrm{SFR_{UV}}\,(M_\odot\, \mathrm{yr^{-1})=7.8\times10^{-29}\ }L_{UV}\,\mathrm{(erg\ s^{-1}\,Hz^{-1})}.
\end{equation}
The calculated SFR values are consistent with the SFR given by {\tt CIGALE} and are adopted in the following analysis. 
We compare the average age computed as $M_*/\mathrm{SFR_{UV}}$ to the burst age fitted by {\tt CIGALE}. We find that these two ages are consistent in general, with mean values of 17 and 13 Myr, respectively. 
This indicates a dominant burst contribution to the total stellar mass in our sample.
Table~\ref{tab:properties} lists the calculated $\mathrm{SFR_{UV}}$ and average ages.  

We also measure the restframe EWs of the three most prominent lines that fall in the NIRCam observations, \oiii, H$\alpha$ and \hb. 
These EW values are based on the best fitted models. For sources in our sample, the \oiii\ and \hb\ lines are redshifted into F200W, and H$\,\alpha$ into the F277W band. As shown in Figure~\ref{fig:sed}, the flux excesses due to line-emission in these two bands are well estimated by the fitted model. 
The results show that EW(\oiii+\hb)s of our sample have a range of 740--6500$\,$\AA, 
with a median of 1700$\,$\AA\ and EW(H$\alpha$)s have a range of 370--2600$\,$\AA\ with a median of 950$\,$\AA. These high EWs qualify our sample as one of the most extreme emission line galaxies (EELGs). 
Correspondingly, over 50\% of the F200W flux and 30\% of the F277W flux can be attributed to these line contributions.

In summary, the SED fitting results reveals the extremely young, dust-poor, and low-mass nature of our sample, which also exhibits extreme emission lines.

\begin{deluxetable}{clr}
\tablewidth{\linewidth} 
\tablecaption{Parameter Space of the {\tt CIGALE} Fitting \label{tab:cigale_para}}
\tablehead{
\colhead{No.} & \colhead{Parameter}  & \colhead{Range}
}
\startdata
(1) & $\tau_\mathrm{{del}}$ & $50-5000$ Myr \\
(2) & $age_\mathrm{{del}}$ & $100-1500$ Myr \\
(3) & $age_\mathrm{{burst}}$ & $1-50$ Myr \\
(4) & $r_\mathrm{{SFR}}$ & $1-10000$ \\
\\
(5) & $Z_*$ & 0.004 \\
(6) & $\log\,U$ & $-2.2--1.2$ \\
(7) & $Z_\mathrm{{gas}}$ & $0.004,\ 0.008$ \\
(8) & $f_\mathrm{{esc}}$ & $0.2-0.7$ \\
\\
(9) & $E(B-V)_{l}$ & $0-0.3$ \\
\enddata
\tablecomments{(1),(2) The e-folding time and age of the delayed population in star formation history. (3) Age of the constant burst. (4) Ratio of the SFR after/before burst. (5) Metallicity of stars. (6) Ionisation parameter. (7) Metallicity of gas. (8) Escape fraction of Lyman continuum photons. (9) The color excess of the nebular lines, converting to the color excess of stellar continuum by multiplying a factor of 0.44.
}
\end{deluxetable}

\begin{deluxetable*}{lCCCCCCc}
\centerwidetable
\tablecaption{Properties of LAE Sample at $z\simeq3.1$\label{tab:properties}}
\tablehead{
\colhead{ID}  & \colhead{$\beta$} & \colhead{$\beta_{0}$} & \colhead{$\log(M_*/M_\odot)$} & \colhead{$\mathrm{SFR_{UV}}$}  &  \colhead{Age}  & \colhead{EW(\oiii+\hb)} & \colhead{EW(H$\alpha$)}\\
\colhead{}  &  \colhead{}   &  \colhead{} & \colhead{} &  \colhead{($M_\odot\,\mathrm{yr^{-1}}$)}  &  \colhead{(Myr)} &  \colhead{($10^3\,$\AA)} &  \colhead{($10^3\,$\AA)}\\
\colhead{(1)}  &  \colhead{(2)}  &  \colhead{(3)}  &  \colhead{(4)}  & \colhead{(5)} & \colhead{(6)} & \colhead{(7)}& \colhead{(8)} 
}
\startdata
UDS-022a 	& -2.76\pm0.10 	& -2.95 & 7.31\pm0.17 	& 2.6 	& 8 	& 3.7 	& 2.3 \\
UDS-022b 	& -2.73\pm0.12 	& -2.90 & 7.17\pm0.18 	& 1.5 	& 10 	& 4.3 	& 2.6 \\
UDS-101 	& -2.42\pm0.08 	& -2.72 & 7.58\pm0.15 	& 2.6 	& 14 	& 2.1 	& 0.9 \\
UDS-140a 	& -2.69\pm0.14 	& -3.07 & 7.05\pm0.18 	& 1.3 	& 8 	& 1.6 	& 1.1 \\
UDS-140b 	& -2.34\pm0.15 	& -2.47 & 7.57\pm0.18 	& 1.0 	& 36 	& 0.8 	& 0.3 \\
UDS-141a 	& -2.78\pm0.09 	& -3.03 & 7.72\pm0.19 	& 7.2 	& 7 	& 2.0 	& 1.0 \\
UDS-141b 	& -2.48\pm0.18 	& -2.69 & 7.25\pm0.24 	& 0.5 	& 34 	& 2.7 	& 0.9 \\
UDS-143 	& -2.32\pm0.05 	& -2.67 & 8.20\pm0.09 	& 7.0 	& 23 	& 0.8 	& 0.4 \\
UDS-144 	& -2.16\pm0.03 	& -2.68 & 8.42\pm0.08 	& 10.7 	& 24 	& 1.0 	& 0.5 \\
UDS-146 	& -2.48\pm0.06 	& -2.81 & 7.74\pm0.15 	& 3.9 	& 14 	& 1.7 	& 1.0 \\
UDS-149 	& -2.41\pm0.06 	& -2.74 & 8.01\pm0.13 	& 6.4 	& 16 	& 1.2 	& 0.7 \\
UDS-152 	& -2.39\pm0.13 	& -2.78 & 8.12\pm0.17 	& 5.8 	& 23 	& 0.7 	& 0.4 \\
UDS-154 	& -2.74\pm0.08 	& -2.78 & 7.40\pm0.13 	& 2.6 	& 10 	& 6.5 	& 2.5 \\
\enddata
\tablecomments{
(1) Object ID. (2)(3)(4) Observed, intrinsic UV slope and stellar mass derived from SED fitting. (5) Uncorrected SFR converted from 1500--2800\AA\ luminosity by equation~\ref{eq:sfr}. (6) Age defined as $M_*/\mathrm{SFR_{UV}}$. (7)(8) Restframe EWs calculated in the best fitted model.
}
\end{deluxetable*}

\section{Results} \label{sec:results}

\begin{figure} %[t]
\centering
\includegraphics[width=0.46\textwidth,trim=20 10 40 40,clip]{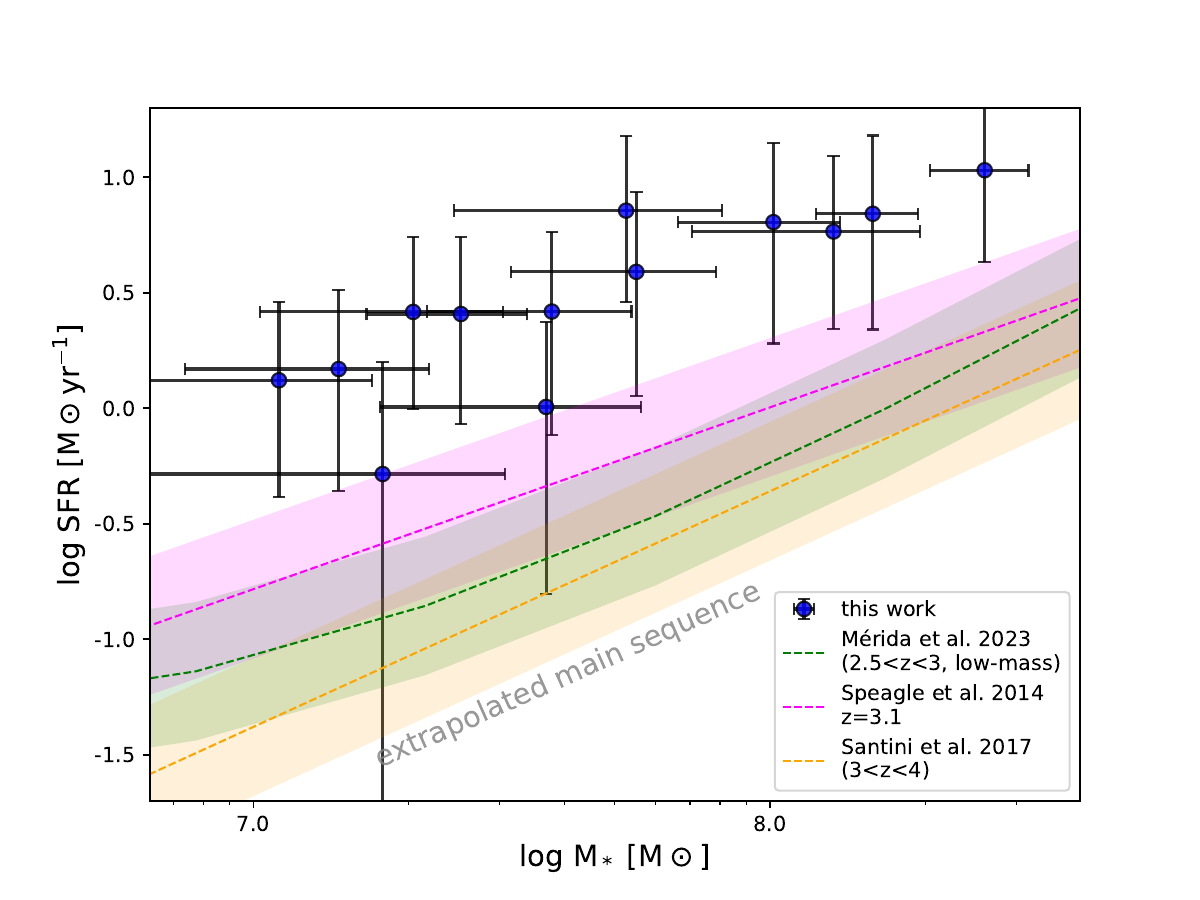}
\caption{SFR to stellar mass distribution of our LAE sample, compared with calibrations of the star-forming main sequence from the literature (S14, S17, M23), extrapolated down to our low-mass range of interest. We use $z=3.1$ to calculate the age of the Universe when plotting the relation given by S14, and the relations from S17 and M23 are selected to be in the redshift bins closest to $z=3.1$. All relations are adjusted to Chabrier IMF scale. 
A uniform 0.3-dex scatter is indicated by the shaded regions around each dashed line. 
The SFR uncertainties of our data points are calculated by adding in quadrature the photometric errors in the restframe UV band plus and an additional 0.25$\,$dex to account for systematic uncertainty in the UV-to-SFR conversion.
\label{fig:mass-sfr}}
\end{figure}

\subsection{Main Sequence}\label{subsec:main_seq}
In Figure~\ref{fig:mass-sfr} we show the SFR to stellar mass distribution of our sample and compare to the extrapolated star-forming main sequence (MS) given by previous works \citep[][hereafter S14, S17 and M23, respectively]{speagle14, santini17, merida23}. All of our sources lie above the three given relations with excesses between 0.3--1.5$\,$dex. 
Here, it is important to emphasize that all three MS relations shown were established on the basis of galaxy samples that do not probe as far down as the low-mass ($\lesssim 10^{8.5}\,M_\odot$) regime explored here.  By lack of further constraints, and acknowledging that the selection effects inherent to our sample inhibit its use for fitting of a MS relation at the low-mass end, we thus simply assume that the same functional form (e.g., linear in the case of S14 and S17) holds down to $10^7$--$10^{8.5} \,M_\odot$ as a description of the typical SF activity in a mass-complete sample of dwarf galaxies.  With this ansatz, we compare our sample with the MS to see whether our LAEs, as a type of star-forming galaxies, lie on or above the MS, which is an unsettled argument.  As we already alluded to above, all sources in our sample appear as starbursting outliers residing well above the MS. 
Taking the MS from S14 as a reference (the one that has the highest SFR in our mass range), our sample is about 0.7$\,$dex higher on average.
There is thus no doubt that our sources feature unusually high specific SFRs compared to the bulk of star-forming galaxies.
However, it is too premature to decide that the whole \lya-emitting population necessarily lies above the MS. Our sample is strongly biased to relatively bright UV continua (which leads to high SFR) through the LAE candidate narrow-band selection and prioritization for subsequent spectroscopic observation. 

Before the advent of JWST, observations of LAEs with higher masses than our sample 
have shown that they either lie on \citep[e.g.,][]{kusakabe15,khusanova20} 
or above \citep[e.g.,][]{hagen16,santos20} 
the star-forming MS. 
Our sample of low-mass, dwarf LAEs lies significantly above the 
extrapolation of the MS based on higher mass galaxies. 
We suspect that either the extrapolation is highly uncertain, 
making it hard to evaluate the genuine MS at the low-mass end (see also Section~\ref{sec:discussion}),
or the dwarf LAEs behave differently from their more massive counterparts. 
For instance, they may be in a similar starburst stage as higher mass LAEs (i.e. similar SFR),
but much earlier in their evolution with little prior star formation, hence making the lower stellar mass the main factor that pushes them above the MS \citep[e.g.,][]{goovaerts24}.

\begin{figure} %[t]
\centering
\includegraphics[width=0.46\textwidth,trim=20 10 40 40,clip]{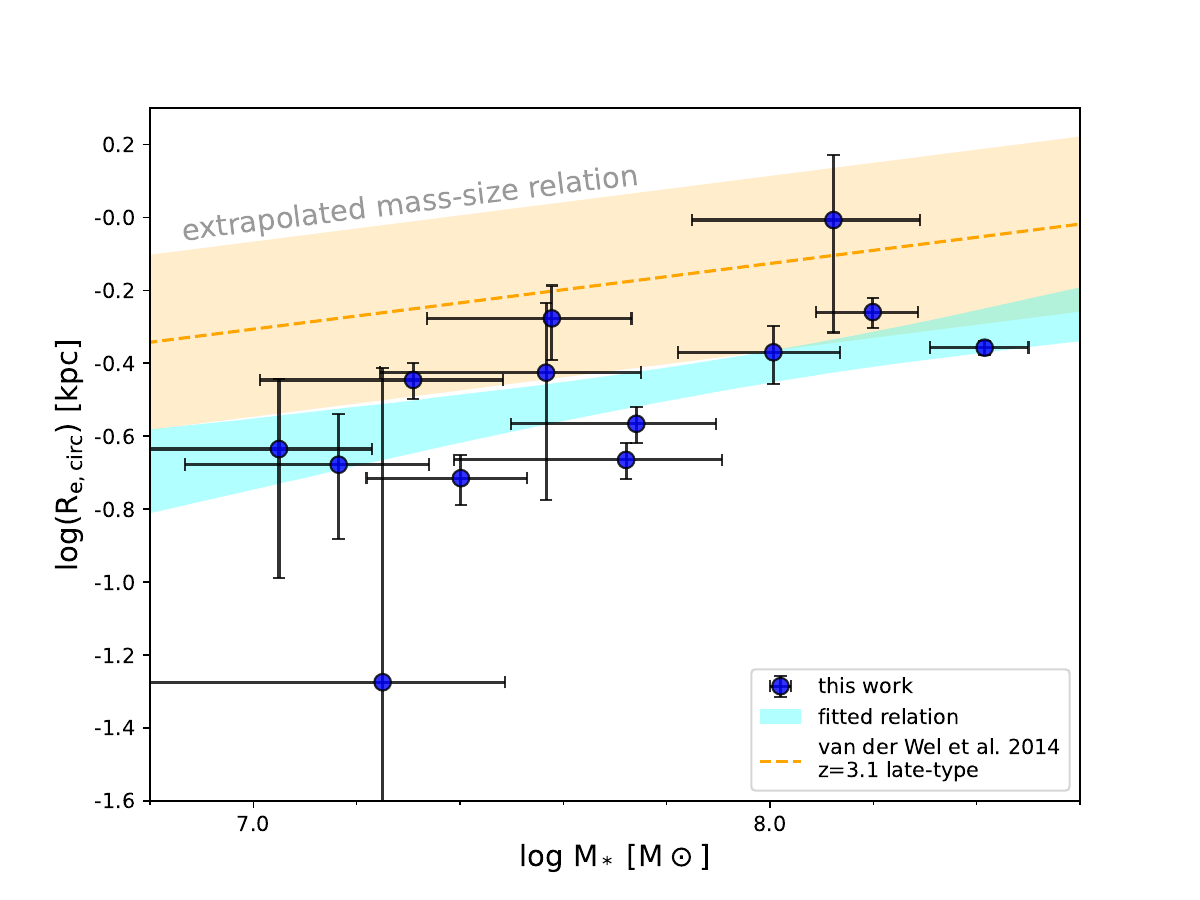}
\caption{Mass-size relation of our LAE sample, with comparison to the \citetalias{vanderwel14} relation for late-type galaxies at $z=3.1$ extrapolated to our mass range. 
The cyan region represents the linear fitted relation within the 16, 84th percentiles derived from MCMC chains. 
Our fitting is conducted with the form of $\log\,R_e=\alpha\,\log\,(M_*/10^7\,M_\odot)+\log\,$A
, and the fitted values of ($\log\,$A, $\alpha$) are ($-0.65^{+0.10}_{-0.10}$, $0.24^{+0.08}_{-0.10}$). 
The orange dashed line in figure represents the extrapolated relation of \citetalias{vanderwel14}, with ($\log\,$A, $\alpha$)=$(-0.31, 0.18)$. A scatter equals to $0.24\,$dex for $R_{e}$ is shown by orange shadow region.
Our sample, although significantly smaller in size and mass from \citetalias{vanderwel14}, has a similar slope between $M_*$ and $R_e$,
indicating similar mass-size relation still holds for less massive star forming galaxies at higher redshift. 
\label{fig:mass-size}}
\end{figure}

\subsection{Mass-Size Relation}\label{subsec:mass-size}
We show the mass-size relation of our $z\approx3.1$ LAEs in Figure~\ref{fig:mass-size} based on the effective radii deduced from the F200W (i.e., rest-optical) images (\S\ref{subsec:size}) and the stellar masses given by SED fitting (\S\ref{subsec:sed}). 
In analogy to \S\ref{subsec:main_seq}, we here also contrast our sample to the extrapolation of a scaling relation from the literature, established on the basis of more massive galaxies. Specifically, we adopt the relation provided by \citet[][hereafter V14]{vanderwel14} at the highest redshift bin (centering at $z=2.75$ with 0.5 as width) and extroplate it to $z=3.1$ by considering the size evolution as a function of redshift using the formula, which is also provided by V14,  constructed in the the lowest mass bin ($10^9$--$10^{9.5}\,M_\odot$). For consistency of the comparison, we convert the \citetalias{vanderwel14} size--mass relation from its original definition in terms of semi-major axis size to circularised $R_e$ by scaling with a median $\rm\sqrt{b/a}$, where the median b/a appropriate for high-z late-type galaxies is 0.54.\footnote{Technically, the CANDELS/3D-HST sample exploited by \citetalias{vanderwel14} reaches up to $z=3$, where its constraints are based on galaxies with masses $\gtrsim 10^{9.5}\,M_\odot$.} 
Our sample occupies a significantly less massive regime (by $\sim$ 3 dex), but additionally also much smaller galaxy sizes, even when accounting for the mass--size relation as established by \citetalias{vanderwel14} (on average $\sim 0.35$ dex smaller). 
We carry out a linear regression of the form $\log\,R_e=\alpha\,\log\,(M_*/10^7\,M_\odot)+\log\,$A on our sample using {\tt emcee} \citep{Foreman-Mackey2013} and obtain best-fitting values for ($\log\,$A, $\alpha$) and the corresponding uncertainties within the 68\% confidence interval of ($-0.65^{+0.10}_{-0.10}$, $0.24^{+0.08}_{-0.10}$).  We note that, while the zero point is substantially lower than that of the size--mass relation by \citetalias{vanderwel14}, the best-fitting slope is consistent.

Other studies on LAEs and EELGs likewise report small sizes, and a correlation between strong \lya\ emission and small physical size is also alluded. (e.g., \citealt{tang21} finds that in an extreme-optical-line-selected sample at $z\simeq2$--3, galaxies with higher EW(\lya) have a slightly smaller size ($R_e=0.49\,$kpc) comparing to galaxies with weaker \lya\ ($R_e=0.76\,$kpc); \citealt{kim21} reports a very compact size of GPs with a typical radius of 0.33$\,$kpc and an anticorrelation between $R_e$ and EW(\lya).)

The size--mass relation for our LAEs has a consistent slope compared to earlier works 
at higher masses \citep[e.g.,][]{vanderwel14}, 
but lies below the extrapolation of the scaling relation towards the low-mass end.  
Whether these dwarf LAEs are indeed more compact than other galaxies of similar mass 
at the same epoch, 
and whether they will remain compact as their masses increase, remain open questions.
Given the lack of statistical samples at low masses ($\log M_*/M_{\odot}<8$),
it is still too early to come to a definite conclusion. 
There are a few factors that affect our interpretation of the observed results. 
Firstly, a single power-law approximation of the size--mass relation may not be appropriate. 
For instance, broken power-laws have been observed in higher-mass samples 
at lower redshift \citep{shen03, vanderwel14, kawinwanichakij21},
yet a steeper slope at the low mass end is also possible although 
earlier studies found the opposite.  
Secondly, the future evolution of LAEs does not need to follow the slope of the size-mass relation 
at the current observed epoch.  
That is, the evolutionary track through
($M_*$, $R_e$) space followed by individual galaxies 
may differ from the trend-line of their distribution at a given epoch, 
as also the zero point of the size-mass relation evolves (i.e., increases) with time. 
Using respectively observations and simulations, both \citetalias{vanderwel14} and \citet{genel18} argued for steeper evolutionary size growth tracks for late-type galaxies than the slope of the size--mass relation, 
which features little evolution with redshift). 
Both star formation and merging could contribute to such steep size growth. 
Disentangling the two requires contrasting galaxy sizes measured in a stellar tracer and an instantaneous star formation tracer.  

For example, 
for a sample of massive ($\gtrsim 10^{9.5}\ M_{\odot}$) star-forming galaxies at $0.7 \lesssim z \lesssim 2.7$, 
\citet{Wilman2020}  used continuum flux as a stellar tracer and H$\alpha$ from integral-field spectroscopy as a tracer for instantaneous star formation, 
and concluded from their relative sizes that 
star formation driven size growth is sufficient only to grow along the size-mass relation, 
not to account for its zero-point evolution.
Processes contributing to such growth in size include star formation and merging.
Mergers, or preferential quenching of the more compact star-forming galaxies, 
would be required to explain the evolution of the scaling relation.  
In the low-mass regime we consider here, 
quenching is anticipated to impact the population negligibly, 
leaving either a strong inside-out trend of star formation or merging as potential routes to bring our starbursting dwarfs onto the regular size--mass relation 
by the time they grow to $\gtrsim 10^9\,M_\odot$.  
Alternatively, they will remain among the more compact objects also at later epochs \citep{vandokkum15, barro14, barro17}.

In principle, size biases may contribute to rendering the measured sizes as compact as observed.  As alluded to in \S\ref{subsec:size}, the F200W and F277W size measurements may not be ideal as a probe of the galaxies' stellar extent, with high-EW line contributions in the median adding up to 50\% and 30\% of the broadband flux, respectively. Enhanced excitation levels in the galaxy centres could boost the central \oiii\ (and consequently F200W) emission, thus skewing the observed sizes to smaller values. In the case of F277W, the main line contribution comes from H$\alpha$, a tracer of newly formed stars. The F150W filter at $z = 3.1$ samples partly below the Balmer break, and may thus likewise receive more weight from younger stellar populations. As a counter argument, however, we raise the conclusion drawn from our SED modelling (\S\ref{subsec:sed}), namely that no evidence is found for an underlying old population. This inference is uniquely enabled by NIRCam's longer wavelength coverage. Indeed, where the S/N allows, consistently small size measurements are obtained in the longer wavelength NIRCam images, albeit increasingly of a marginally resolved nature.

A more in depth assessment of size growth scenarios from the dwarf to normal galaxy regime will require the collection of more data on the general distribution of galaxy sizes down to masses of around $10^7$--$10^8\,M_\odot$.

\subsection{UV continuum}\label{subsec:uvslope}
\begin{figure} %[t]
\centering
\includegraphics[width=0.46\textwidth,trim=20 10 40 40]{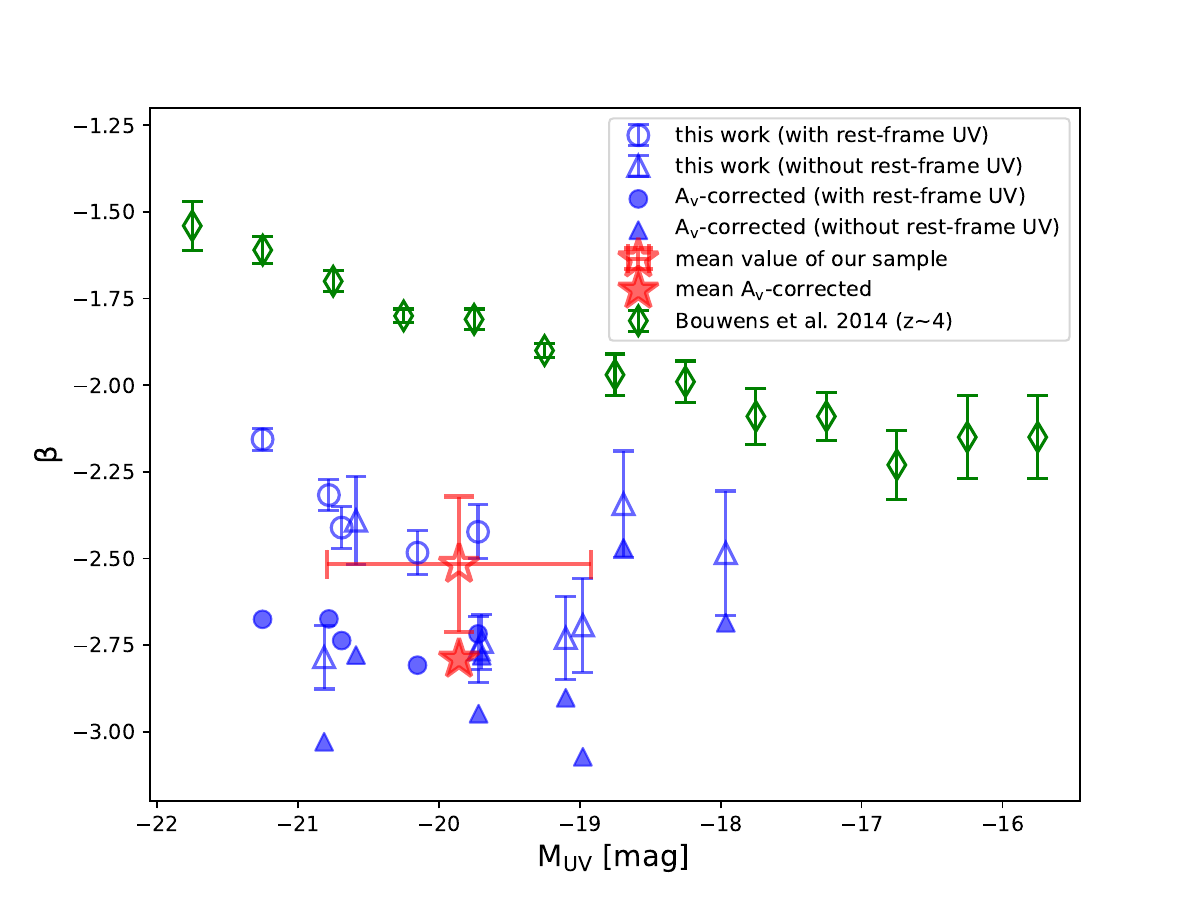}
\caption{
UV slope $\beta$ as a function of absolute UV magnitude ${M_{UV}}$. 
The filled symbols represent the extinction corrected $\beta_0$, 
while the open symbols are the uncorrected data.
With circles, we mark the sources in our sample whose SEDs are fitted with extra restframe-UV bands (see \S\ref{subsec:sed}), which means that their $\beta$ values are empirically constrained. Sources without photometry restriction in restframe-UV bands are marked with triangles. The red star represents the average position of our whole sample, while its error bars show the standard deviation in two dimensions.
The binned $M_{UV}$--$\beta$ relation of a sample of UV-selected dropout galaxies by \citet{Bouwens2014} is shown for reference in green diamonds.
\label{fig:uvslope}}
\end{figure}

In Figure~\ref{fig:uvslope} we consider the relation between the absolute UV magnitude ($M_{UV}$) and the UV slope ($\beta$) of LAEs in our sample.
Specifically, we quantify $\beta$ on the best-fitting stellar population model with (filled symbols) and without (open symbols) extinction correction. 
Extinction could be one of the dominating factors that affect the UV slopes \citep{wilkins11},
but as discussed in Section \ref{subsec:sed}, 
all of our targets have little to no dust extinction. 
As mentioned in \S\ref{subsec:sed}, only 5 objects in our sample are modeled with additional restframe UV photometries, which constrains the $\beta$ measurement directly. For other sources, their $\beta$ values are implied by the stellar population model fitted at longer wavelength. We show the results of both fitting modes in Figure~\ref{fig:uvslope}, but with different symbols.
For reference, we also show the binned $M_{UV}$--$\beta$ relation for high-z galaxies (at $z \sim 4$) compiled by \citet{Bouwens2014}.  Their sample comprises UV-selected galaxies identified via the Lyman-break (a.k.a. dropout) technique, drawn from the deepest rest-UV imaging campaigns with HST (XDF, HUDF), and thus extends down to fainter UV magnitudes. 
In comparison, our sample occupies the luminous side of the parameter space, but of significantly steeper slopes, indicating negligible reddening of the UV continuum. 

We see no reason why the offset in UV slopes (particularly in this direction, towards bluer colours) could be attributed to the modest ($\Delta z \approx 0.9$) offset in redshift between the two samples.  If anything, one may expect the typical galaxy at a given UV luminosity to have matured slightly, in terms of its ISM composition, over the intervening half billion years. The latter would constitute an offset towards redder UV slopes.  Instead, our LAE selected objects (which turn out to share rest-optical line characteristics in common with EELGs) should be taken as representing the tail population of youngest systems with no/little dust obscuration with respect to the underlying UV-selected galaxy population.  Indeed, similar differences between the LAE and LBG populations have previously been reported by \citet{stark10}.

\section{Discussion} \label{sec:discussion}
From studies in the nearby Universe, and from galaxy formation simulations, dwarf galaxies are known to experience bursty star formation histories (SFHs).  For this reason, the existence of a population of MS outliers should not necessarily come as a surprise.  While the various selection steps, from NB selection to spectroscopic targeting, do not guarantee that we are complete, even to a starbursting population defined by a fixed $\Delta {\rm MS}$ threshold, we can contrast the size of our selected LAE sample to the number of MS outliers naively expected given a set of simplified conditions.  Specifically, for this back-of-the-envelope estimation, we assume that, at $z = 3.1$: \\
1) The galaxy stellar mass function established by \citet{weaver22} can be extrapolated down to the low-mass ($10^7$--$10^{8.5}\,M_\odot$) regime of consideration here. \\
2) The MS relation by S14 holds down to such low masses. \\
3) The distribution of SFR values around the MS is unimodal and has a constant (i.e., mass-independent) Gaussian scatter of 0.3$\,$dex.

Under these assumptions, and taking into account the exploited survey volume ($\sim 150\,{\rm arcmin}^2$ for the overlap between NB imaging and presently released PRIMER coverage, and $z = 3.05$--$3.16$ based on the NB selection), we estimate there would be 19 galaxies with $\Delta {\rm MS} > 0.7$ dex over the $10^7$--$10^{8.5}\,M_\odot$ mass range (i.e., overlapping with the parameter range of our sample).  In other words, the 13 sources in our sample, corresponding to a co-moving number density of $3.5\times10^{-4}\,{\rm {Mpc}^{-3}}$, could be accommodated under the above conditions.  Not all galaxies feature significant amounts of escaping \lya\ photons though. \citet{sarkar19} compare the UV luminosity function of Lyman-break galaxies (LBGs) by \citet{Reddy2008} to the UV luminosity function of LAEs by \citet{ouchi08}, and estimate a \lya\ fraction of $\sim 10\%$ at $z=3.1$, with relatively little dependence on UV luminosity across the dynamic range probed. If we assume that the same \lya\ fraction applies down to our mass regime and uniformly across all $\Delta {\rm MS}$ (both potentially questionable assumptions), one would require an enhanced MS scatter, of $\gtrsim 0.5\,$dex, to account for the 13 observed LAEs (and more so if other high-$\Delta {\rm MS}$ outliers exist but were missed by the NB selection or from spectroscopic targeting).

The above numbers were obtained adopting the extrapolated MS from S14.  Were we to adopt extrapolations of the MS calibration by S17 or M23 instead, which has SFR values dropping more steeply with decreasing mass, the requirement on an increased gaussian scatter of the MS at the low-mass end (or equivalently, a super-gaussian tail towards high $\Delta {\rm MS}$) would become correspondingly more pressing.

We conclude that circumstancial arguments can be made to hint at an enhanced MS scatter in the dwarf galaxy regime, but increased number statistics, a well-defined selection function and improved insight in the validity of the assumed steps in our reasoning are needed to make a firmer case for a more bursty nature of low-mass galaxies, relative to their more massive counterparts at the same epoch.

\section{Summary} \label{sec:summary}
In this paper, we present our first results on a sample of spectroscopically confirmed LAEs at $z\approx3.1$, observed in the rest-optical to NIR wavelengths by JWST/NIRCam. With released data from the PRIMER program, we zoom in on 10 `LAE positions' observed in 8 NIR bands, through which we study their morphology and stellar properties. 
Overall, our sample is composed of galaxies with very similar properties in every aspect, including a small size (median $R_e=0.36\,$kpc), very blue UV continuum (median $\beta=-2.48$) with little to no dust extinction, extreme optical line emission (median ${\rm EW_{rest}}$(\oiii+\hb)$=1700$\AA), low mass ($\log\,(M_*/M_\odot)=7.1$--$8.4$),
and high specific SFR ($0.3$--$1.4\times10^{-7}\,{\rm yr}^{-1}$).

Among the 10 positions, three are resolved into pair-like compositions with separations $<0.9''$. We deem that the sources in pairs are plausibly related to each other because they all have the same feature in their SEDs, namely a F200W excess contributed by extremely strong \oiii+\hb\ lines redshifted into this filter, as do other sources in our sample. However, neither the narrow-band images nor the spectroscopy allow determining with certainty whether the \lya\ emission is from both objects or only one of them. In this work, we treat them as individual LAEs. Besides possible pairs, asymmetrical structures are also captured in three other sources. Some hints of widening or double-peaked features can be seen in the \lya\ spectra of these asymmetrical and pair-like sources. 

Using effective radii measured on the F200W images, stellar masses derived by SED fitting, and SFRs from $1500$--$2800$\AA\ model fluxes, we compare the mass-size relation of our sample to the extrapolated empirical one (Figure~\ref{fig:mass-size}) and their mass-SFR distribution to the extrapolated main sequence (Figure~\ref{fig:mass-sfr}). In both diagrams, our sample shows large offsets from the extrapolated empirical relations. Specifically, our selected LAEs are on average $\sim0.7\,$dex higher in SFR than the MS, and $\sim0.35\,$dex smaller in size than the ridge of the mass-size relation. Also in terms of their inferred UV slope $\beta$, the LAEs stand out with respect to the underlying population of UV-selected (dropout) galaxies by their bluer colours.

Considering their numbers, we speculate that this starbursting dwarf galaxy population hints at an increased MS scatter or more pronounced tail of above-MS outliers at the low-mass end, although increased number statistics from ongoing and upcoming JWST programs are needed to place such constraints more firmly.

\begin{acknowledgments}
This work is sponsored by the National Key R\&D Program of China for grant No.\ 2022YFA1605300, 
the National Nature Science Foundation of China (NSFC) grants No.\ 12273051 and\ 11933003. 
Support for this work is also partly provided by the CASSACA.  
The authors gratefully acknowledge support from the Royal Society International Exchanges Scheme (IES\textbackslash R1\textbackslash 211140) and the Chinese Academy of Sciences President's International
Fellowship Initiative (grant no. 2022VMB0004).

This work is based on observations made with the NASA/ESA/CSA James Webb Space Telescope. The data were obtained from the Mikulski Archive for Space Telescopes at the Space Telescope Science Institute, which is operated by the Association of Universities for Research in Astronomy, Inc., under NASA contract NAS 5-03127 for JWST. These observations are associated with program GO-1837. The authors acknowledge the PRIMER team for developing their observing program with a zero-exclusive-access period.
\end{acknowledgments}

\vspace{5mm}
\facilities{JWST (NIRCam)}

\software{{\tt astropy} \citep{astropy}, {\tt GALFIT}, {\tt CIGALE}, {\tt emcee}, {\tt matplotlib}} 
%          Cloudy \citep{2013RMxAA..49..137F}, 
%          Source Extractor \citep{1996A&AS..117..393B}
%          }

\bibliography{reference}{}
\bibliographystyle{aasjournal}
\end{CJK*}
\end{document}